\documentclass{article}

\usepackage{microtype}
\usepackage{graphicx}
\usepackage{subfigure}
\usepackage{booktabs} 
\usepackage{amsfonts}
\usepackage{tkz-graph}
\usepackage{hyperref}
\usepackage{amsthm}
\usetikzlibrary {positioning}
\newtheorem{defn}{Definition}

\usepackage[accepted]{icml2019}

\icmltitlerunning{Anomaly Detection with Joint Representation Learning of Content and Connection}

\begin{document}

\twocolumn[
\icmltitle{Anomaly Detection with Joint Representation Learning of Content and Connection}




\begin{icmlauthorlist}
\icmlauthor{Junhao Wang*}{mcgill}
\icmlauthor{Renhao Wang*}{ubc}
\icmlauthor{Aayushi Kulshrestha}{mcgill}
\icmlauthor{Reihaneh Rabbany}{mcgill}
\end{icmlauthorlist}

\icmlaffiliation{mcgill}{School of Computer Science, McGill University, Montreal, Canada}
\icmlaffiliation{ubc}{Department of Computer Science, University of British Columbia, Vancouver, Canada}

\icmlcorrespondingauthor{Junhao Wang}{junhao@umich.edu}

\icmlkeywords{Machine Learning, ICML}

\vskip 0.3in
]



\printAffiliationsAndNotice{\icmlEqualContribution} 

\begin{abstract}

Social media sites are becoming a key factor in politics. These platforms are easy to manipulate for the purpose of distorting information space to confuse and distract voters. Past works to identify disruptive patterns are mostly focused on analyzing the content of tweets. In this study, we jointly embed the information from both user posted content as well as a user's follower network, to detect groups of densely connected users in an unsupervised fashion. We then investigate these  dense sub-blocks of users to flag anomalous behavior. In our experiments, we study the tweets related to the upcoming 2019 Canadian Elections, and observe a set of densely-connected users engaging in local politics in different provinces, and exhibiting troll-like behavior.

\end{abstract}

\section{Introduction}
There have been multiple recent studies on how social networks have been manipulated to foster divisions among people, and the tactics used for information operations. 
\citet{wilson2018assembling} define Information Operations as the suite of methods used to influence others through the dissemination of propaganda and disinformation. The aim is to paralyze the decision making abilities of individuals, thus making the public vulnerable. With these operations at large, democracy stands threatened. A case in point is the 2016 US Presidential Elections which are believed to have been swayed through social media by interference from Russian trolls and bots \cite{badawy2018analyzing}. This was further validated by Twitter in 2018, when they reported possible engagement of 1.4 billion users with suspected ``trolls" from the Russian government funded Internet Research Agency \cite{policy2018update}.

Twitter is one of the most commonly used platforms to mobilize public at the time of political unrest. Therefore, it is imperative that we develop tools to identify Information Operations at an early stage, leading to a healthy democratic society. 
Towards this goal, we analyze the activity on Twitter related to the upcoming elections in Canada, to proactively monitor the interactions on this platform. The main contributions of the current work are:

\begin{figure}[t]
\begin{center}
\centerline{\includegraphics[width=\columnwidth]{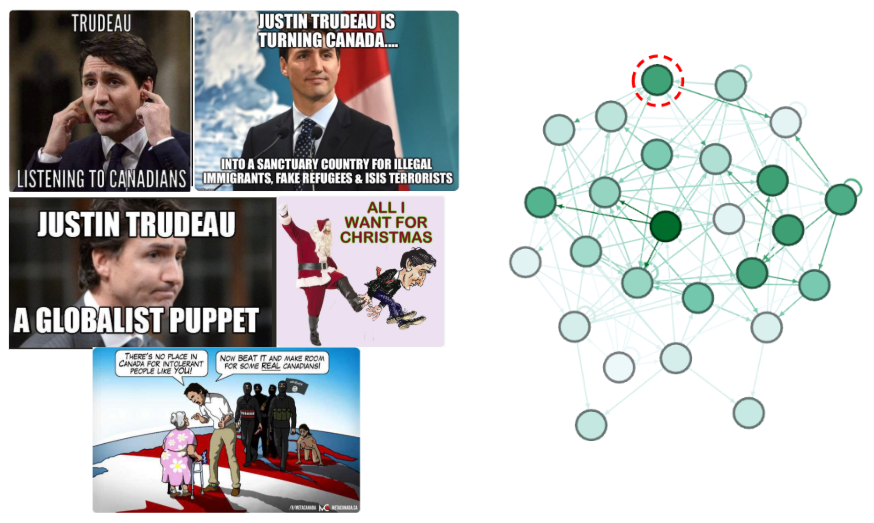}}
\caption{Sample Memes used by Suspicious Anomalous User in Cluster \#8 and its Follower Network Location}
\label{anomaly}
\end{center}
\end{figure}
\begin{itemize}
    \item We create a joint autoencoder based solution to the problem by formulating Information Operation detection as dense sub-block detection on binary attributed graph, which encodes both the content of tweets and the connections in the Twitter follower network. The dense sub-blocks are detected  using density-based clustering on learned node embeddings of the graph.
    \item We design an adaptive hyperparameter selection method by generating task-specific synthetic data, thus solving the problem of lack of objective evaluation standard for this unsupervised anomaly detection tasks.
    \item We demonstrate the application of our solution to real-world data by identifying a sub-block of suspicious and tightly connected users, as well as a suspicious account exhibiting behaviors related to Information Operations. 
\end{itemize}

\section{Data Collection}
In this paper, our main focus would be to study the political interactions of Twitter users surrounding Canadian 2019 Federal Election, and identify groups of users that seek to disseminate crafted messages (propaganda). 
The dataset comprises of 38,498 tweets from 7,298 distinct Twitter users. The tweets were collected using the Twitter Streaming API, using the following hashtags -  \#Trudeau, \#TrudeauMustGo, \#cdnpoli, \#TrudeauResign, \#LavScam, \#SNCgate, \#StandWithTrudeau. We also collected the list of followers for each of the 7,298 users in our dataset to construct a follower network, resulting in a total of 474,459 connections.
The hashtags are further formulated as a vector of size 3,047 to represent the attributes of each node or user, denoting whether the user used a certain hashtag in the dataset. We represent the entire data as concatenated adjacency and attribute matrix as shown in Figure \ref{aja-attr}.






\section{Method}



\begin{defn}[\textbf{Binary Attributed Graph}]\label{def:bag}
A binary attributed graph $\mathcal{G} = (\mathcal{V},\mathcal{E},\mathbf{X})$ consists of: (1) the set of nodes $\mathcal{V}=\{v_1,v_2,\dots,v_n\}$, where $|\mathcal{V}| = n$; (2) the set of edges $\mathcal{E}$, where $|\mathcal{E}| = m$; and (3) the binary node attribute matrix $\mathbf{X}\in\{0,1\}^{n\times d}$, where the $i^{th}$ row vector $\mathbf{x}_i\in \{0,1\}^d, (i=1\dots n)$ is the binary attribute information for the $i^{th}$ node.
\end{defn}

\begin{defn}[\textbf{Binary Attributed Graph Dense Sub-block}]\label{def:dense}
For a binary attributed graph $\mathcal{G} = (\mathcal{V},\mathcal{E},\mathbf{X})$ and a density threshold $t \in [0,1]$, $t$-induced dense sub-blocks are set of tuples $\{(\mathcal{S_V},\mathcal{S_\mathbf{X}})\}$ where $\mathcal{S_V}$ is subset of nodes $\mathcal{V}$ and $\mathcal{S_\mathbf{X}}$ is subset of binary attributes of $\mathbf{X}$, such that: (1) subgraph induced by $\mathcal{S_V}:\mathcal{S_V}(\mathcal{V},\mathcal{E})$ has network density above $t$, and (2) bipartite graph induced by top nodes $\mathcal{S_V}$, bottom nodes $\mathcal{S_\mathbf{X}}$ and edges $\mathcal{E}_b$, has network density above $t$, where $\mathcal{E}_b$ correspond to activation of binary attributes. We denote adjacency matrix of $\mathcal{G}$ by $\mathbf{A}$, attribute matrix by $\mathbf{X}$.
\end{defn}

We represent our data as a Binary Attributed Graph with adjacency matrix $\mathbf{A}$ encoding follower relations and binary attribute matrix $\mathbf{X}$ encoding user hashtag usage. We are interested in detecting binary attributed graph dense sub-block. 
%
%
Our approach to tracking Twitter political interactions and potentially identifying Information Operations is a 3-fold process: (1) learn node embeddings using joint autoencoder to minimize reconstruction error of the adjacency matrix and attribute matrix and preserve pairwise Jaccard distance of input vectors, (2) apply density-based clustering on the node embeddings, (3) conduct manual inspection of filtered clusters.


\subsection{Joint Autoencoder}
Our joint autoencoder architecture is inspired by \cite{8594881}, where we extend their loss functions to deal with joint information of attribute and adjacency matrix.
\begin{defn}[Joint Autoencoder]
A joint autoencoder shown in Figure \ref{encoder} is characterized by $(\phi^e_\mathbf{A}, \phi^e_\mathbf{X}, \phi^e_\mathbf{J},\phi^d_\mathbf{A},\phi^d_\mathbf{X})$ where $\phi$ can be a function represented by a single layer of neural network or composition of multiple layers, $\phi^e$ is encoding function and $\phi^d$ is decoding function. Subscripts of $\phi :$ $\mathbf{A}, \mathbf{X}, \mathbf{J}$ denote the information $\phi^e$ encodes from or $\phi^d$ decodes into, where $\mathbf{A}$ and $\mathbf{X}$ are adjacency and attribute matrix of $\mathcal{G}$ defined previously, and $\mathbf{J}$ is concatenated latent representation of them: $concat(\phi^e_\mathbf{A}(\mathbf{A}), \phi^e_\mathbf{X}(\mathbf{X}))$. Decoders $\phi^d_\mathbf{A}$ and $\phi^d_\mathbf{X}$ transform joint latent representation $\mathbf{J}$ to approximation of $\mathbf{A}, \mathbf{X}: \hat{\mathbf{A}}, \hat{\mathbf{X}}$.
\end{defn}

The joint reconstruction error weighted by hyperparameters $w_\mathbf{A}$ and $w_\mathbf{X}$, with attention weights $\mathbf{W}_{att}^{\mathbf{A}}$ and $\mathbf{W}_{att}^{\mathbf{X}}$ is calculated by

\begin{equation}
    \mathbf{H} = \phi^e_{\mathbf{J}}(\mathbf{J})
\end{equation}
\begin{equation}
    \mathcal{L}_{recon}^{\mathbf{A}} = || (\phi^d_{\mathbf{A}}(\mathbf{H}) - \mathbf{A}) \odot  \mathbf{W}_{att}^{\mathbf{A}}||_F^2
\end{equation}
\begin{equation}
 \mathcal{L}_{recon}^{\mathbf{X}} = || (\phi^d_{\mathbf{X}}(\mathbf{H}) - \mathbf{X}) \odot \mathbf{W}_{att}^{\mathbf{X}} ||_F^2
\end{equation}
\begin{equation}
 \mathcal{L}_{recon} = w_{\mathbf{A}} \mathcal{L}_{recon}^{\mathbf{A}} + w_{\mathbf{X}} \mathcal{L}_{recon}^{\mathbf{X}}
\end{equation}

Besides reconstruction loss, we define similarity loss as the discrepancy between pairwise Euclidean distance of $\mathbf{H}$ and pairwise Jaccard distance of $\mathbf{A}$ and $\mathbf{X}$, weighted by the same $w_{\mathbf{A}}$ and $w_{\mathbf{X}}$. In order to compare these 2 different distance metrics, we apply a logit transformation on the pairwise Euclidean distance to compress its range to $[0,1]$, the same as the range of pairwise Jaccard distance. Let $\mathbf{S}_{Jar}^{\mathbf{X}}$ be the pairwise Jaccard distance of rows of $\mathbf{X}$, similarly $\mathbf{S}_{Jar}^{\mathbf{A}}$ for $\mathbf{A}$, and $\mathbf{S}_{Euc}^{\mathbf{H}}$ be the pairwise Euclidean distance for latent vectors $\mathbf{H}$, and choose $\lambda \geq 0$:

\begin{equation}
    \mathcal{L}_{sim}^{\mathbf{A}} = || exp(-\lambda \mathbf{S}_{Euc}^{\mathbf{H}}) - \mathbf{S}_{Jar}^{\mathbf{A}} ||_F^2
\end{equation}
\begin{equation}
    \mathcal{L}_{sim}^{\mathbf{X}} = || exp(-\lambda \mathbf{S}_{Euc}^{\mathbf{H}}) - \mathbf{S}_{Jar}^{\mathbf{X}} ||_F^2
\end{equation}
\begin{equation}
 \mathcal{L}_{sim} = w_{\mathbf{A}} \mathcal{L}_{sim}^{\mathbf{A}} + w_{\mathbf{X}} \mathcal{L}_{sim}^{\mathbf{X}}
\end{equation}

The joint loss to minimize is weighted combination of reconstruction loss and similarity loss with weights $w_{recon}$ and $w_{sim}$, plus L2 regularization loss at every layer:

\begin{equation}
    \mathcal{L}_{joint} = w_{recon} \mathcal{L}_{recon} + w_{sim} \mathcal{L}_{sim} + \mathcal{L}_{reg}
\end{equation}

In practice, we train on sampled batches instead of the entire data matrix. For each epoch, we select node $v_i\in\mathcal{V}$ uniformly at random, and sample set of nodes $\{v_j:v_j\in \{\mathcal{V}-v_i\}\}$ according to some distribution $D$ related to the similarity between $v_i$ and $v_j$. 


\subsection{Density-based Clustering}

We transform the problem of dense sub-block detection of the binary attributed graph to density-based clustering of latent embeddings of nodes in Euclidean space, thus enabling the use of established density-based clustering algorithm such as DBSCAN \citep{ester1996density}, as well as making the process more interpretable. We first apply dimensionality reduction using Uniform Manifold Approximation and Projection \citep{mcinnes2018umap} on $\mathbf{H}$ to get $\mathbf{H}_{reduced}$ with 2 dimensions, and then apply DBSCAN on $\mathbf{H}_{reduced}$ with parameters that cater to the size of the clusters that we are interested to study. Then we define dense clusters as clusters that induce subgraphs of Twitter follower network whose network density is above a specified threshold. In practice, we look at the top $k$ densest clusters returned by DBSCAN. The motivation for using network density to indicate anomaly is related to \citet{wilson2018assembling}'s definition of Information Operations: we aim to study densely connected users exhibiting similar hashtag usage, thus potentially showing similar attitudes towards certain events or ideologies, which is indicative of the presence of Information Operation.


\section{Experiments}

We first conduct synthetic experiments to choose the best set of hyperparameters for exploratory analysis on real data, as well as to demonstrate the effectiveness of our algorithm on binary attributed graph dense sub-block detection compared to FRAUDAR \cite{hooi2016fraudar}, a classical baseline for dense sub-block detection with only adjacency matrices, and DOMINANT \cite{ding2019deep}, a Graph Convolutional Network (GCN) based approach that utilizes both adjacency and attribute matrices. We then create a joint ``fingerprint" of identified clusters based on both the graph topology of cluster-induced subgraph, and attributes of nodes in the cluster, which could potentially be used to identify Information Operations in Canadian 2019 Federal Election. We also manually inspect the nodes in the three clusters with highest cluster-induced network density, and find some suspicious accounts that might have engaged in Information Operations.

\subsection{Hyper-Parameter Tuning}

\begin{figure}[!htb]
\begin{center}
\centerline{\includegraphics[width=.8 \columnwidth]{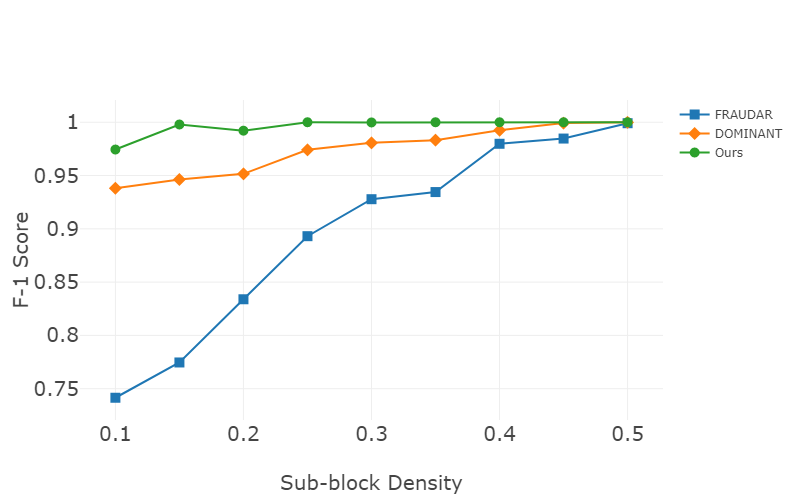}}
\caption{Synthetic Experiment Performance}
\label{syn-perform}
\end{center}
\end{figure}

We inject artificial dense sub-block anomalies into our Twitter data in order to tune our algorithm to perform well for the unsupervised anomaly detection task. 
With the injected data, we conduct a random search of the hyperparameter space and identify the best hyperparameter option by F-1 score, with labels being anomaly or non-anomaly. Then we use it to identify interesting dense clusters on the real data without dense sub-block injection. We show that our method outperforms both baselines across all injected sub-block densities in Figure \ref{syn-perform}.


\subsubsection{Synthetic Data Generation}
For adjacency matrix, we inject dense subgraph by injecting random dense graph with a specified density and size at sub-block indices. For attribute matrix entries, we create an empirical distribution of hashtag usage indicating how likely a random person from a sub-block would use certain hashtags, and apply add-$k$ smoothing on this empirical distribution. Next, we sharpen the distribution by applying an exponential factor to it: $exp(\lambda \ \cdot)$ where $\lambda$ controls for how concentrated the transformed distribution is. By sampling a certain number of hashtags from this distribution, we simulate the presence of Information Operations, where a group of highly connected users tweet a subset of hashtags. Finally we inject the bipartite graph with the specified density at these sub-block and attribute indices. For our experiment, we inject 3 dense sub-blocks of size 500, and use the same network density for both adjacency matrix and attribute matrix, from $0.1$ to $0.5$ with $0.05$ interval.





\begin{figure*}[th!]
\vspace{-.1in}
\begin{center}
\centerline{\includegraphics[height=0.55\columnwidth,width=2.0\columnwidth,trim={0cm 1cm 0 2cm},clip]{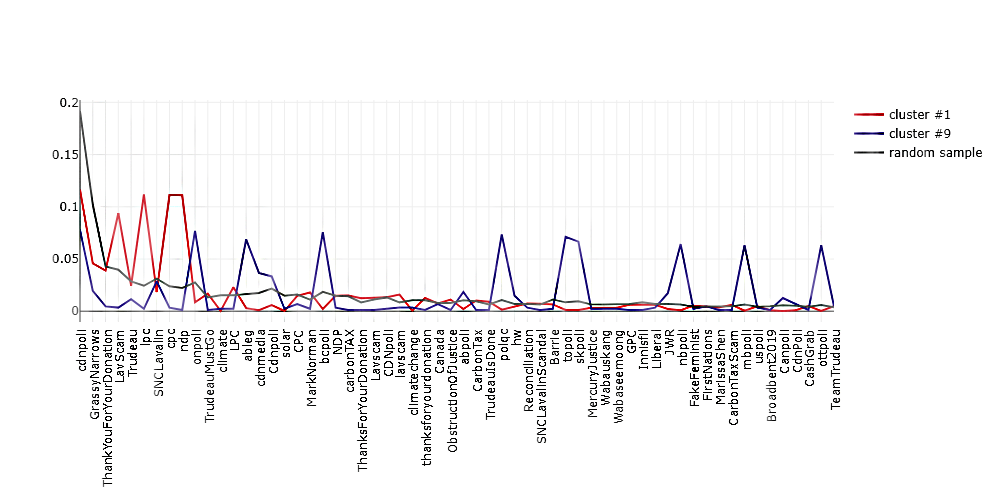}}
\caption{Hashtag Fingerprint: Per-cluster relative usage frequency for popular hashtags}
\label{hashfinger}
\end{center}
\vskip -0.2in
\end{figure*}
\subsection{Results}
Using the best hyperparameter option, we create 10 clusters. 
For each cluster and its corresponding induced sub-graph of follower network, we generate hashtag fingerprint, which reflects user attribute information, as well as clustering fingerprint, which reflects key network topology information. We put our focus on 2 of the densest clusters  (\#9, \#1), and report interesting exploratory findings.




For each cluster, we define hashtag fingerprint as the relative usage frequency of popular hashtags within cluster. Note that usage here refers to whether a hashtag is used or not in the dataset for a given user, and frequency refers to the number of users in a cluster using certain hashtag. A high relative usage frequency corresponds to highly used hashtag in a cluster. Hashtag fingerprints for cluster \#9 and \#1, and a randomly sampled set of users are shown in Figure \ref{hashfinger}. Similarly, for each cluster-induced subgraph of the follower network, we define clustering fingerprint (shown in Figure \ref{clustfinger}) by the probability density of node clustering coefficients in the subgraph. Clustering coefficient of each user captures the connected-ness of its neighbors.

By analyzing the hashtag fingerprint we note that cluster \#9 exhibits interesting spikes on hashtags related to diverse locations: Alberta (\#ableg), British Columbia (\#bcpoli), Quebec (\#polqc), Toronto (\#topoli), Saskatchewan (\#skpoli), New Brunswick (\#nbpoli), Manitoba (\#mbpoli) and Ottawa (\#ottpoli). This is counter-intuitive; we normally assume people engage with each other locally, but we clearly see multi-regional cluster of users in close contact with each other. 

For cluster \#1, the hashtag usage centers around a recently heated political scandal related to government and corporate corruption (\#LavScam), and a few prominant political parties: the Liberal Party of Canada (\#lpc), the Conservative Party of Canada (\#cpc), and the New Democratic Party (\#ndp). This reflects the fact that an emergent cluster of users related to different political parties are talking about the recent scandal.

The hashtag fingerprint for both clusters identified through our algorithm reveals interesting insights that would otherwise be hard to obtain by going through the tweets manually. On the other hand, the hashtag fingerprint of a random sample is highly centered around the most popular hashtags and cannot yield much insight into the user group.

Inspecting clustering fingerprints in Figure \ref{clustfinger}, both cluster \#1 and \#9 exhibit a more spread-out distribution compared to random sample. In particular, cluster \#9 where tweet hashtags are related to diverse locations exhibits clustering coefficient distribution centered around value ($0.1$ to $0.2$) higher than what we would expect for such a multi-regional cluster. To investigate the reason for such phenomena is an interesting direction of future empirical study. 

\subsubsection{Sample User}

We finally manually inspect the Twitter profile page of users in top 3 densest clusters (\#8, \#9, \#1), and for each cluster, we visualized its cluster-induced subgraph and per-node HITS authority score \citep{kleinberg1999authoritative}. We use darker green gradient to denote nodes with higher HITS authority score.

Shown in Figure \ref{anomaly}, we identify one Twitter account highlighted with a dotted red line that exhibits behaviors suspicious of Information Operations. The suspicious user account was created in August 2017. Since December 2017, the user started consistently creating and spreading divisive tweets and memes that demote Justin Trudeau and his administration. A sample of the political memes deployed by this user is shown in Figure \ref{anomaly}. Furthermore, this user changed it user handle twice from mid-April to mid-May. Such high frequency of changing user handles might be related to malicious intent \cite{Jain:2016:DUC:2888451.2888452}. Both the identified user's posted content and behavior are suspicious of engaging in Information Operations.

\section{Conclusion}

In this work, we proposed an embedding based solution to track Twitter political interactions and potentially identify anomalous user groups. This approach is built on the recent advances on representation learning for graphs and hence provides a scalable solution for this domain. We contribute three key insights: (1) Information Operations detection on Twitter can be formulated as a dense sub-block detection problem on binary attributed graphs that encode both network topology and user attribute information; (2) dense sub-block detection can be formulated as a density-based clustering problem on latent Euclidean embeddings of the nodes in the graph; (3) each cluster identified by the algorithm can be assigned a joint fingerprint that yields insight which is otherwise not possible through manual inspection of tweets.  For code and more information see \url{https://sites.google.com/view/jointdetect/}



\newpage
\nocite{langley00}
\bibliography{example_paper}

\begin{thebibliography}{10}
\providecommand{\natexlab}[1]{#1}
\providecommand{\url}[1]{\texttt{#1}}
\expandafter\ifx\csname urlstyle\endcsname\relax
  \providecommand{\doi}[1]{doi: #1}\else
  \providecommand{\doi}{doi: \begingroup \urlstyle{rm}\Url}\fi

\bibitem[Badawy et~al.(2018)Badawy, Ferrara, and Lerman]{badawy2018analyzing}
Badawy, A., Ferrara, E., and Lerman, K.
\newblock Analyzing the digital traces of political manipulation: The 2016
  russian interference twitter campaign.
\newblock In \emph{2018 IEEE/ACM International Conference on Advances in Social
  Networks Analysis and Mining (ASONAM)}, pp.\  258--265. IEEE, 2018.

\bibitem[Ding et~al.(2019)Ding, Li, Bhanushali, and Liu]{ding2019deep}
Ding, K., Li, J., Bhanushali, R., and Liu, H.
\newblock Deep anomaly detection on attributed networks.
\newblock 2019.

\bibitem[Ester et~al.(1996)Ester, Kriegel, Sander, Xu,
  et~al.]{ester1996density}
Ester, M., Kriegel, H.-P., Sander, J., Xu, X., et~al.
\newblock A density-based algorithm for discovering clusters in large spatial
  databases with noise.
\newblock In \emph{Kdd}, volume~96, pp.\  226--231, 1996.

\bibitem[Hooi et~al.(2016)Hooi, Song, Beutel, Shah, Shin, and
  Faloutsos]{hooi2016fraudar}
Hooi, B., Song, H.~A., Beutel, A., Shah, N., Shin, K., and Faloutsos, C.
\newblock Fraudar: Bounding graph fraud in the face of camouflage.
\newblock In \emph{Proceedings of the 22nd ACM SIGKDD International Conference
  on Knowledge Discovery and Data Mining}, pp.\  895--904. ACM, 2016.

\bibitem[Jain \& Kumaraguru(2016)Jain and
  Kumaraguru]{Jain:2016:DUC:2888451.2888452}
Jain, P. and Kumaraguru, P.
\newblock On the dynamics of username changing behavior on twitter.
\newblock In \emph{Proceedings of the 3rd IKDD Conference on Data Science,
  2016}, CODS '16, pp.\  6:1--6:6, New York, NY, USA, 2016. ACM.
\newblock ISBN 978-1-4503-4217-9.
\newblock \doi{10.1145/2888451.2888452}.
\newblock URL
  \url{http://doi.acm.org.proxy3.library.mcgill.ca/10.1145/2888451.2888452}.

\bibitem[Kleinberg(1999)]{kleinberg1999authoritative}
Kleinberg, J.~M.
\newblock Authoritative sources in a hyperlinked environment.
\newblock \emph{Journal of the ACM (JACM)}, 46\penalty0 (5):\penalty0 604--632,
  1999.

\bibitem[McInnes et~al.(2018)McInnes, Healy, and Melville]{mcinnes2018umap}
McInnes, L., Healy, J., and Melville, J.
\newblock Umap: Uniform manifold approximation and projection for dimension
  reduction.
\newblock \emph{arXiv preprint arXiv:1802.03426}, 2018.

\bibitem[Policy(2018)]{policy2018update}
Policy, T.~P.
\newblock Update on twitter’s review of the 2016 us election.
\newblock \emph{Retrieved April}, 15:\penalty0 2018, 2018.

\bibitem[{Wang} et~al.(2018){Wang}, {Zhou}, {Wu}, {Dang}, {Zhu}, and
  {Wang}]{8594881}
{Wang}, H., {Zhou}, C., {Wu}, J., {Dang}, W., {Zhu}, X., and {Wang}, J.
\newblock Deep structure learning for fraud detection.
\newblock In \emph{2018 IEEE International Conference on Data Mining (ICDM)},
  pp.\  567--576, Nov 2018.
\newblock \doi{10.1109/ICDM.2018.00072}.

\bibitem[Wilson et~al.(2018)Wilson, Zhou, and Starbird]{wilson2018assembling}
Wilson, T., Zhou, K., and Starbird, K.
\newblock Assembling strategic narratives: Information operations as
  collaborative work within an online community.
\newblock \emph{Proceedings of the ACM on Human-Computer Interaction},
  2\penalty0 (CSCW):\penalty0 183, 2018.

\end{thebibliography}
\bibliographystyle{icml2019}
\newpage
\appendix
\section{Plots and Images}

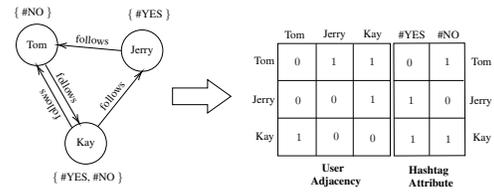
\begin{figure}[!htb]
    \centering
    \resizebox{.8 \columnwidth}{!}{
    \tikzset{every picture/.style={line width=0.75pt}} 

\begin{tikzpicture}[x=0.75pt,y=0.75pt,yscale=-1,xscale=1]

\draw   (385.33,63.62) -- (520.06,63.62) -- (520.06,198.35) -- (385.33,198.35) -- cycle ;
\draw   (609.26,63.1) -- (609.78,198.05) -- (524.72,198.38) -- (524.2,63.42) -- cycle ;
\draw    (385.33,108.04) -- (520.06,108.04) ;

\draw    (385.33,153.11) -- (520.06,153.11) ;

\draw    (524.43,153.11) -- (610.19,153.11) ;

\draw    (524.1,108.04) -- (608.18,108.04) ;

\draw    (431.27,63.65) -- (431.27,198.18) ;

\draw    (475.67,64.32) -- (475.67,198.85) ;

\draw    (566.99,63.48) -- (566.99,198) ;

\draw   (71.52,65.23) .. controls (71.52,51.52) and (82.64,40.4) .. (96.36,40.4) .. controls (110.07,40.4) and (121.19,51.52) .. (121.19,65.23) .. controls (121.19,78.95) and (110.07,90.06) .. (96.36,90.06) .. controls (82.64,90.06) and (71.52,78.95) .. (71.52,65.23) -- cycle ;
\draw   (197.68,72.18) .. controls (197.68,58.47) and (208.79,47.35) .. (222.51,47.35) .. controls (236.22,47.35) and (247.34,58.47) .. (247.34,72.18) .. controls (247.34,85.9) and (236.22,97.02) .. (222.51,97.02) .. controls (208.79,97.02) and (197.68,85.9) .. (197.68,72.18) -- cycle ;
\draw   (130.13,183.44) .. controls (130.13,169.72) and (141.25,158.6) .. (154.96,158.6) .. controls (168.68,158.6) and (179.8,169.72) .. (179.8,183.44) .. controls (179.8,197.15) and (168.68,208.27) .. (154.96,208.27) .. controls (141.25,208.27) and (130.13,197.15) .. (130.13,183.44) -- cycle ;
\draw    (197.68,72.18) -- (123.18,65.41) ;
\draw [shift={(121.19,65.23)}, rotate = 365.19] [color={rgb, 255:red, 0; green, 0; blue, 0 }  ][line width=0.75]    (10.93,-3.29) .. controls (6.95,-1.4) and (3.31,-0.3) .. (0,0) .. controls (3.31,0.3) and (6.95,1.4) .. (10.93,3.29)   ;

\draw    (138.37,164.56) -- (97.34,91.81) ;
\draw [shift={(96.36,90.06)}, rotate = 420.58000000000004] [color={rgb, 255:red, 0; green, 0; blue, 0 }  ][line width=0.75]    (10.93,-3.29) .. controls (6.95,-1.4) and (3.31,-0.3) .. (0,0) .. controls (3.31,0.3) and (6.95,1.4) .. (10.93,3.29)   ;

\draw    (169.17,163.57) -- (221.26,98.58) ;
\draw [shift={(222.51,97.02)}, rotate = 488.71] [color={rgb, 255:red, 0; green, 0; blue, 0 }  ][line width=0.75]    (10.93,-3.29) .. controls (6.95,-1.4) and (3.31,-0.3) .. (0,0) .. controls (3.31,0.3) and (6.95,1.4) .. (10.93,3.29)   ;

\draw    (106.59,89.07) -- (146.31,157.87) ;
\draw [shift={(147.31,159.6)}, rotate = 240] [color={rgb, 255:red, 0; green, 0; blue, 0 }  ][line width=0.75]    (10.93,-3.29) .. controls (6.95,-1.4) and (3.31,-0.3) .. (0,0) .. controls (3.31,0.3) and (6.95,1.4) .. (10.93,3.29)   ;

\draw   (259,117) -- (301,117) -- (301,107) -- (329,127) -- (301,147) -- (301,137) -- (259,137) -- cycle ;

\draw (454.84,221.78) node  [align=left] {\textbf{ \ \ \ User}\\\textbf{Adjacency}};
\draw (569.01,223.23) node  [align=left] {\textbf{Hashtag}\\\textbf{Attribute}};
\draw (405.51,53.53) node  [align=left] {Tom};
\draw (453.27,53.53) node  [align=left] {Jerry};
\draw (499.01,53.53) node  [align=left] {Kay};
\draw (546.09,53.53) node  [align=left] {\#YES};
\draw (589.81,52.85) node  [align=left] {\#NO};
\draw (452.59,85.14) node   {$1$};
\draw (498.33,85.14) node   {$1$};
\draw (369.19,84.47) node  [align=left] {Tom};
\draw (368.52,132.23) node  [align=left] {Jerry};
\draw (369.86,175.95) node  [align=left] {Kay};
\draw (629.5,85.81) node  [align=left] {Tom};
\draw (628.83,133.57) node  [align=left] {Jerry};
\draw (630.17,177.29) node  [align=left] {Kay};
\draw (409.55,132.23) node   {$0$};
\draw (499.01,130.88) node   {$1$};
\draw (408.87,85.81) node   {$0$};
\draw (452.7,130.98) node   {$0$};
\draw (410.89,175.95) node   {$1$};
\draw (455.29,175.27) node   {$0$};
\draw (499.11,176.72) node   {$0$};
\draw (543.4,86.49) node   {$0$};
\draw (588.47,85.81) node   {$1$};
\draw (545.42,132.23) node   {$1$};
\draw (589.14,131.55) node   {$0$};
\draw (547.44,177.96) node   {$1$};
\draw (589.81,178.64) node   {$1$};
\draw (96.36,65.23) node  [align=left] {Tom};
\draw (222.51,72.18) node  [align=left] {Jerry};
\draw (154.96,183.44) node  [align=left] {Kay};
\draw (231.07,29.15) node  [align=left] {\{ \#YES \}};
\draw (157.44,224.33) node  [align=left] {\{ \#YES, \#NO \}};
\draw (91.7,26.34) node  [align=left] {\{ \#NO \}};
\draw (162.91,59.27) node [rotate=-5.06] [align=left] {follows};
\draw (186.75,125.82) node [rotate=-308.73] [align=left] {follows};
\draw (133.11,116.88) node [rotate=-58.96] [align=left] {follows};
\draw (111.26,136.75) node [rotate=-237.67] [align=left] {follows};

\end{tikzpicture}
    }
    \caption{Concatenated Adjacency and Attribute Matrix}
    \label{aja-attr}
\end{figure}

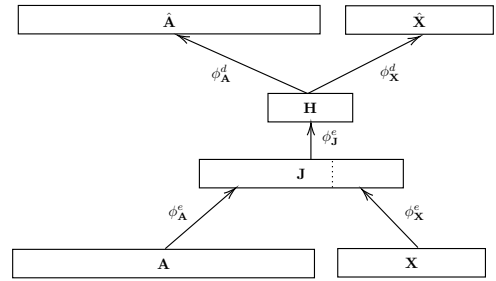
\begin{figure}[!htb]
    \centering
    \resizebox{.8 \columnwidth}{!}{
    \tikzset{every picture/.style={line width=0.75pt}} 

\begin{tikzpicture}[x=0.75pt,y=0.75pt,yscale=-1,xscale=1]

\draw   (117,264.69) -- (385.3,264.69) -- (385.3,290) -- (117,290) -- cycle ;
\draw   (406,263.98) -- (537.3,263.98) -- (537.3,289.3) -- (406,289.3) -- cycle ;
\draw   (344,125.98) -- (419.3,125.98) -- (419.3,151.3) -- (344,151.3) -- cycle ;
\draw   (122,47.69) -- (390.3,47.69) -- (390.3,73) -- (122,73) -- cycle ;
\draw   (413,47.98) -- (544.3,47.98) -- (544.3,73.3) -- (413,73.3) -- cycle ;
\draw    (252.65,264.09) -- (313.79,211.39) ;
\draw [shift={(315.3,210.09)}, rotate = 499.24] [color={rgb, 255:red, 0; green, 0; blue, 0 }  ][line width=0.75]    (10.93,-3.29) .. controls (6.95,-1.4) and (3.31,-0.3) .. (0,0) .. controls (3.31,0.3) and (6.95,1.4) .. (10.93,3.29)   ;

\draw    (477.3,263.09) -- (428.69,212.53) ;
\draw [shift={(427.3,211.09)}, rotate = 406.12] [color={rgb, 255:red, 0; green, 0; blue, 0 }  ][line width=0.75]    (10.93,-3.29) .. controls (6.95,-1.4) and (3.31,-0.3) .. (0,0) .. controls (3.31,0.3) and (6.95,1.4) .. (10.93,3.29)   ;

\draw   (283,184.69) -- (464.3,184.69) -- (464.3,210) -- (283,210) -- cycle ;
\draw  [dash pattern={on 0.84pt off 2.51pt}]  (401.3,186.09) -- (401.3,212.09) ;

\draw    (382.15,184.73) -- (382.15,154.23) ;
\draw [shift={(382.15,152.23)}, rotate = 450] [color={rgb, 255:red, 0; green, 0; blue, 0 }  ][line width=0.75]    (10.93,-3.29) .. controls (6.95,-1.4) and (3.31,-0.3) .. (0,0) .. controls (3.31,0.3) and (6.95,1.4) .. (10.93,3.29)   ;

\draw    (379.3,126.09) -- (265.13,74.9) ;
\draw [shift={(263.3,74.09)}, rotate = 384.15] [color={rgb, 255:red, 0; green, 0; blue, 0 }  ][line width=0.75]    (10.93,-3.29) .. controls (6.95,-1.4) and (3.31,-0.3) .. (0,0) .. controls (3.31,0.3) and (6.95,1.4) .. (10.93,3.29)   ;

\draw    (379.3,126.09) -- (477.52,75.99) ;
\draw [shift={(479.3,75.09)}, rotate = 512.98] [color={rgb, 255:red, 0; green, 0; blue, 0 }  ][line width=0.75]    (10.93,-3.29) .. controls (6.95,-1.4) and (3.31,-0.3) .. (0,0) .. controls (3.31,0.3) and (6.95,1.4) .. (10.93,3.29)   ;

\draw (264.5,232.09) node   {$\phi ^{e}_{\mathbf{A}}$};
\draw (474.5,232.09) node   {$\phi ^{e}_{\mathbf{X}}$};
\draw (251.15,277.34) node   {$\mathbf{A}$};
\draw (471.65,276.64) node   {$\mathbf{X}$};
\draw (373.65,197.34) node   {$\mathbf{J}$};
\draw (381.65,138.64) node   {$\mathbf{H}$};
\draw (399.5,165.09) node   {$\phi ^{e}_{\mathbf{J}}$};
\draw (478.65,60.64) node   {$\hat{\mathbf{X}}$};
\draw (256.15,60.34) node   {$\hat{\mathbf{A}}$};
\draw (302.66,108.01) node   {$\phi ^{d}_{\mathbf{A}}$};
\draw (452.5,107.09) node   {$\phi ^{d}_{\mathbf{X}}$};

\end{tikzpicture}
    }
    \caption{Joint Autoencoder Architecture}
    \label{encoder}
\end{figure}

\begin{figure}[ht]
\begin{center}
\centerline{\includegraphics[width=1.0\columnwidth]{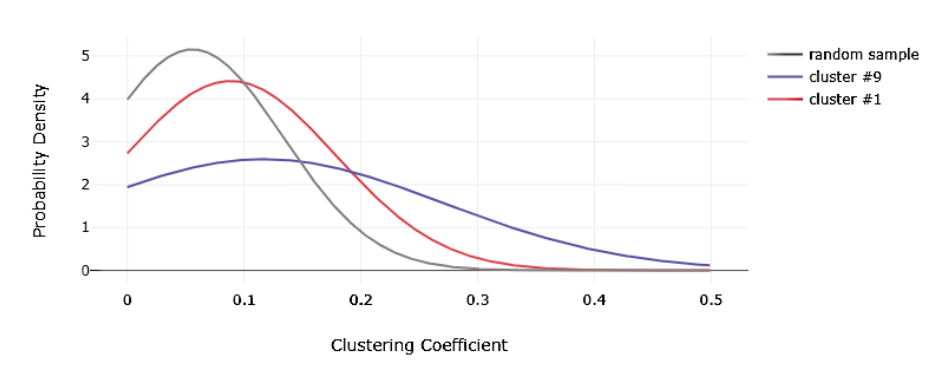}}
\caption{Clustering Fingerprint: Per-cluster probability density for node clustering coefficients}
\label{clustfinger}
\end{center}
\vskip -0.2in
\end{figure}
\end{document}